%% file: MAIN.tex
\documentclass[a4paper,11pt]{article}
\pdfoutput=1 

\usepackage{jinstpub} 

\usepackage[version=3]{mhchem}
\usepackage{siunitx}
\usepackage{natbib}
\usepackage{amsmath} 
\newcommand{\angstrom}{\textup{\AA}}

\title{Test of low radioactive molecular sieves for radon filtration in SF$_6$ gas-based rare-event physics experiments}

\author[a,1]{R.R. Marcelo Gregorio,\note{Corresponding author.}}
\author[a]{N.J.C. Spooner,}
\author[a]{J. Berry,}
\author[a]{A. C. Ezeribe,}
\author[b]{K. Miuchi,}
\author[c]{H. Ogawa}
\author[a]{and A. Scarff}

\affiliation[a]{Department of Physics and Astronomy, University of Sheffield, South Yorkshire, S3 7RH, United Kingdom}
\affiliation[b]{Department of Physics, Graduate School of Science, Kobe University, 1-1 Rokkodai-cho, Nada-ku, Kobe, Hyogo, 657-8501, Japan}
\affiliation[c]{CST Nihon University, Surugadai, Kanda, Chiyoda-ku, Tokyo, 180-0011, Japan}

\emailAdd{robert.gregorio@sheffield.ac.uk}

\abstract{ Type $5\angstrom$ molecular sieves (MS) have been demonstrated to remove radon from SF$_6$ gas. This is important for ultra-sensitive SF$_6$ gas-based directional dark matter and related rare-event physics experiments, as radon can provide a source of unwanted background events. Unfortunately, commercially available sieves intrinsically emanate radon at levels not suitable for ultra-sensitive physics experiments. A method to produce a low radioactive MS has been developed in Nihon University (NU). In this work, we explore the feasibility of the NU-developed $5\angstrom$ type MS for use in such experiments. A comparison with a commercially available Sigma-Aldrich $5\angstrom$ type MS was made. The comparison was done by calculating a parameter indicating the amount of radon intrinsically emanated by the MS per unit radon captured from SF$_6$ gas. The measurements were made using a specially adapted DURRIDGE RAD7 radon detector. The NU-developed $5\angstrom$ MS emanated radon up to 61±9$\%$ less per radon captured (2.1±0.1)$\times 10^{-3}$, compared to the commercial Sigma-Aldrich MS (5.4±0.4)$\times 10^{-3}$, making it a better candidate for use in a radon filtration setup for future ultra-sensitive \ce{SF6} gas based experiments.}


\begin{document}
\maketitle
\flushbottom

\section{Introduction}
\label{sec:intro}
\input{introduction}

\section{Molecular sieve}
\label{sec:MS}
\input{MS}

\newpage 
\section{Radon emanation test}
\label{sec:emanation}
\input{emanation}

\section{Radon filtration test}
\label{sec:filtration}
\input{filtration}

\section{Molecular sieve feasibility}
\label{sec:comparison}
\input{comparison}

\section{Macroscopic geometry optimisation}
\label{sec:geometry}
\input{geometry}

\newpage
\section{Conclusions}
\label{sec:conclusions}
In this work, a method to determine the suitability of a molecular sieve candidate for use in ultra-sensitive SF$_6$ gas-based rare-event physics experiments was proposed. A parameter indicating the amount of radon intrinsically emanated per unit radon captured by the sieve from SF$_6$ was experimentally determined. A low radioactive MS candidate developed by Nihon University in collaboration with Union Showa K.K was tested and compared to a commercially manufactured Sigma-Aldrich MS. It was found that the NU-developed MS emanated  48±15$\%$ less per radon captured (2.8±0.7)$\times 10^{-3}$, compared to the commercial Sigma-Aldrich MS (5.4±0.4)$\times 10^{-3}$. An attempt to improve this result was made by crushing the MS into a powder, increasing its surface area to volume ratio. The powder form significantly increased the radon capture efficiency. However, the intrinsic radon emanation also worsen.  Overall, there was no change in the emanation per radon captured parameter for the NU-developed MS powder (2.1±0.1)$\times 10^{-3}$. The total reduction in our parameter with the powdered NU-developed MS compared to the Sigma-Aldrich MS was 61±9\%, which is within errors of the original granulated NU-developed MS. In practice, when constructing a radon-filtration setup, it is recommended an amount of MS considerably more than required be used. This is because radon capture is a probabilistic kinetic process, using more sieves will reduce the timescale required to reach radon adsorption equilibrium within the gas detector. The NU-developed MS provides a suitable candidate for use in a radon filtration setup for future ultra-sensitive \ce{SF6} gas-based experiments.  However, efforts towards minimising the emanation per radon captured parameter must continue so that the total amount of MS allowed by the radioactive budget of an experiment is maximised.


\acknowledgments
The authors would like to acknowledge support for this work through the EPSRC Industrial CASE with DURRIDGE UK Award grant (EP/R513313/1).




\input{REFERENCES}


\end{document}

%% file: introduction.tex
Minimisation of radon contamination is essential in ultra-sensitive gas rare-event physics experiments including directional dark matter experiments \cite{battat_Drift_2014, miuchi_cygnus_2020}. The decay of radon gas in these experiments can produce unwanted background events, able to mimic genuine signals. Radon contamination comes primarily from detector materials. These materials can contain a trace amount of \ce{^238 U} due to inevitable material contamination. The radon gas is produced in the \ce{^238 U} decay chain. Radon contamination levels are dependent on the size of the experiment. For instance, the radon contamination in the NEWAGE experiment, a small 23 x 28 x 30 cm$^3$ direction dark matter detector, was measured to be 0.2 mBq \cite{Nakamura2014}. Whereas, in the larger DRIFT experiment, a 1.5 x 1.5 x 1.5 m$^3$ gas-based direction dark matter detector, was measured to be 372±66 mBq \cite{battat_radon_2014}. The increase in contamination levels is due to the additional amount of material. Therefore for large-scale plans, radon contamination levels will inevitably be higher. The radon background goal for future large-scale experiments, such as CYGNUS-1000 \cite{Vansen_2020}, which utilises a 1000 m$^3$ detector, is less than $\sim$ 1 mBq. Many gas rare-event physics experiments manage radon contamination with continuous flow and disposal of the target gas.

The gas \ce{SF6} has been identified as a target gas for use in future directional dark matter searches \cite{phan_novel_2017}. However, \ce{SF6} is the most potent greenhouse gas, making the method of continuous flow and disposal problematic \cite{Ezeribe2019}. Furthermore, climate change initiatives, such as the EU F-Gas directive, heavily regulate the use of \ce{SF6} and aim to reduce total use fivefold by 2030, compared with levels in 2014 \cite{anonymous_eu_2016}. This restriction creates a hurdle for ultra-sensitive \ce{SF6}-based rare-event physics experiments, particularly for large-scale plans such as CYGNUS-1000. Therefore, an alternative method to minimise radon contamination must be implemented for future \ce{SF6} based experiments, where the gas is reused and recycled.

The demonstration of radon removal from \ce{SF6} gas was a significant advance towards a radon filtration system \cite{ezeribe_demonstration_2017}. In principle, the \ce{SF6} gas can be continuously recirculated and reused by filtration with the $5\angstrom$ type molecular sieve, reducing the total amount of \ce{SF6} used. In the past, activated charcoal have been studied for radon filtration \cite{Pushkin2018, Abe2012}. However, molecular sieves offer superior gas selectivity due to their specific pore sizes \cite{Rolando2018}.

Unfortunately, commercial molecular sieves intrinsically emanate radon at levels unsuitable for ultra-sensitive rare-event physics experiments. Commercially available sieves are primarily used in the petroleum industry, where having low radioactive content is not essential, so the manufacturer does not need to screen materials for radioactive content in production. Therefore, it can be assumed that any commercially manufactured molecular sieves will not meet the required low radioactivity level.

Recently, a method to produce low radioactive molecular sieves was developed in Nihon University. A $5\angstrom$ type Nihon University developed molecular sieve may provide a suitable candidate for use in a radon filtration setup for ultra-sensitive \ce{SF6} gas based rare-event physics experiments. In this paper, we investigate the feasibility of a $5\angstrom$ type NU-developed molecular sieve for use in a radon filtration setup for ultra-sensitive \ce{SF6} gas based rare-event physics experiments. This was done by measuring the radon emanated from the NU-developed molecular sieve and its radon capture efficiency from \ce{SF6}. These measurements were compared with the commercial, Sigma-Aldrich manufactured, molecular sieve used in the original radon filtration demonstration in 2017 \cite{ezeribe_demonstration_2017}.

%% file: MS.tex
\begin{figure}[ht]
\centering 
\includegraphics[height=4.5cm]{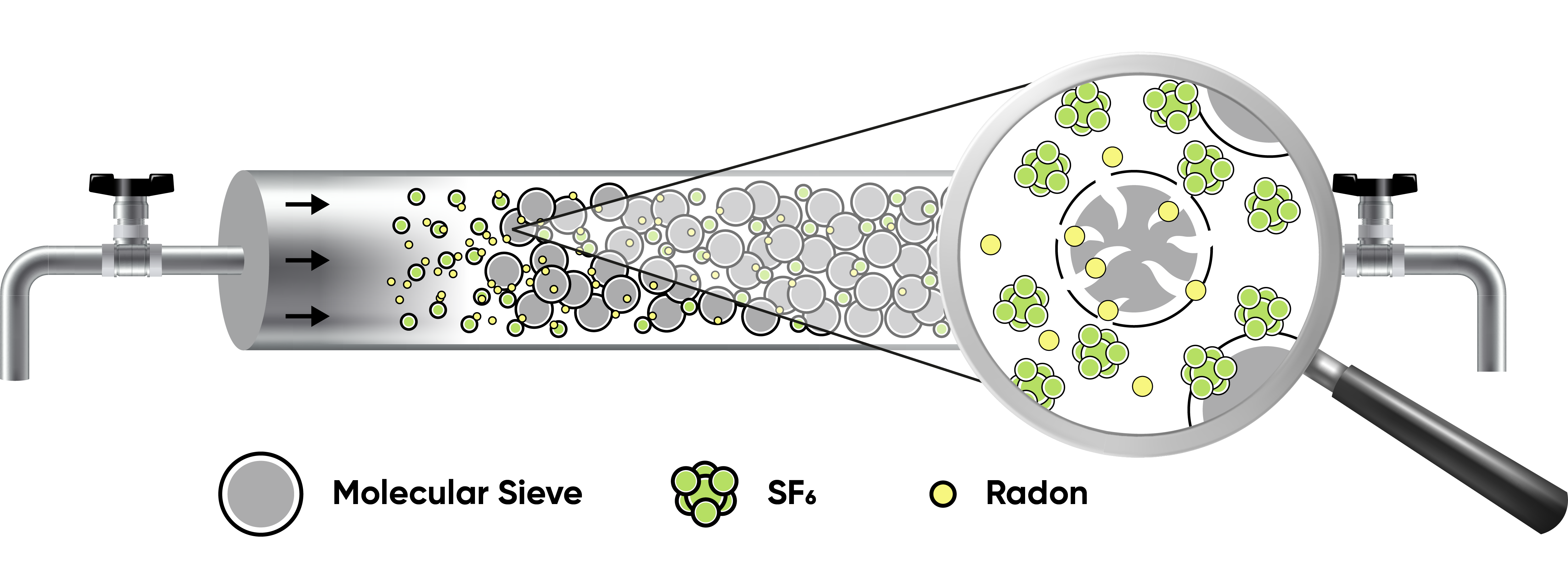}
\caption{\label{fig:MS_FILTER} Illustration of the separation of smaller radon atom from larger \ce{SF6} molecules using molecular sieves. The arrows represent the direction of flow of the gas inside the molecular sieve vessel.}
\end{figure}
\begin{figure}[ht]
\centering 
\includegraphics[height=6cm]{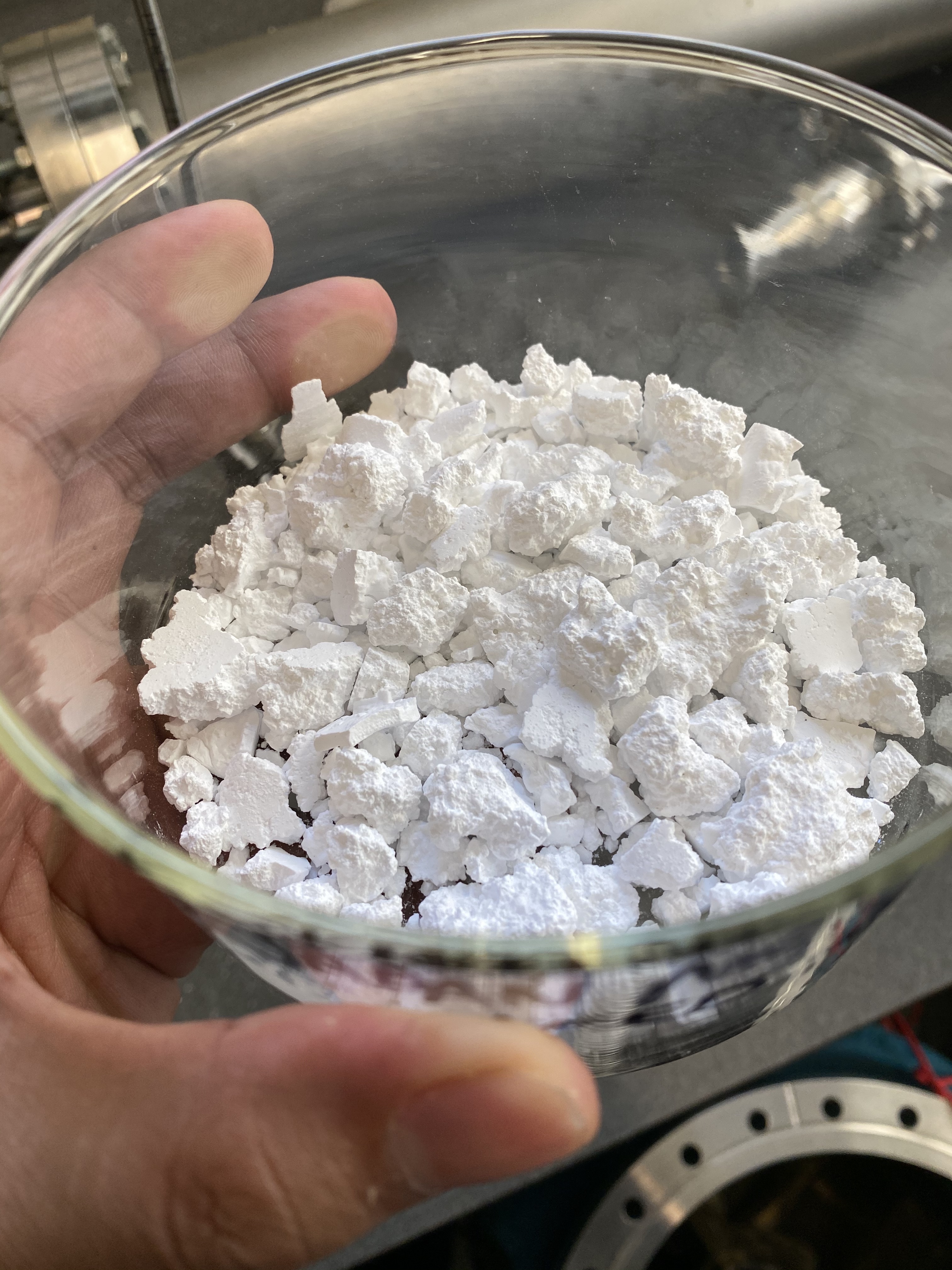}
\qquad
\includegraphics[height=6cm]{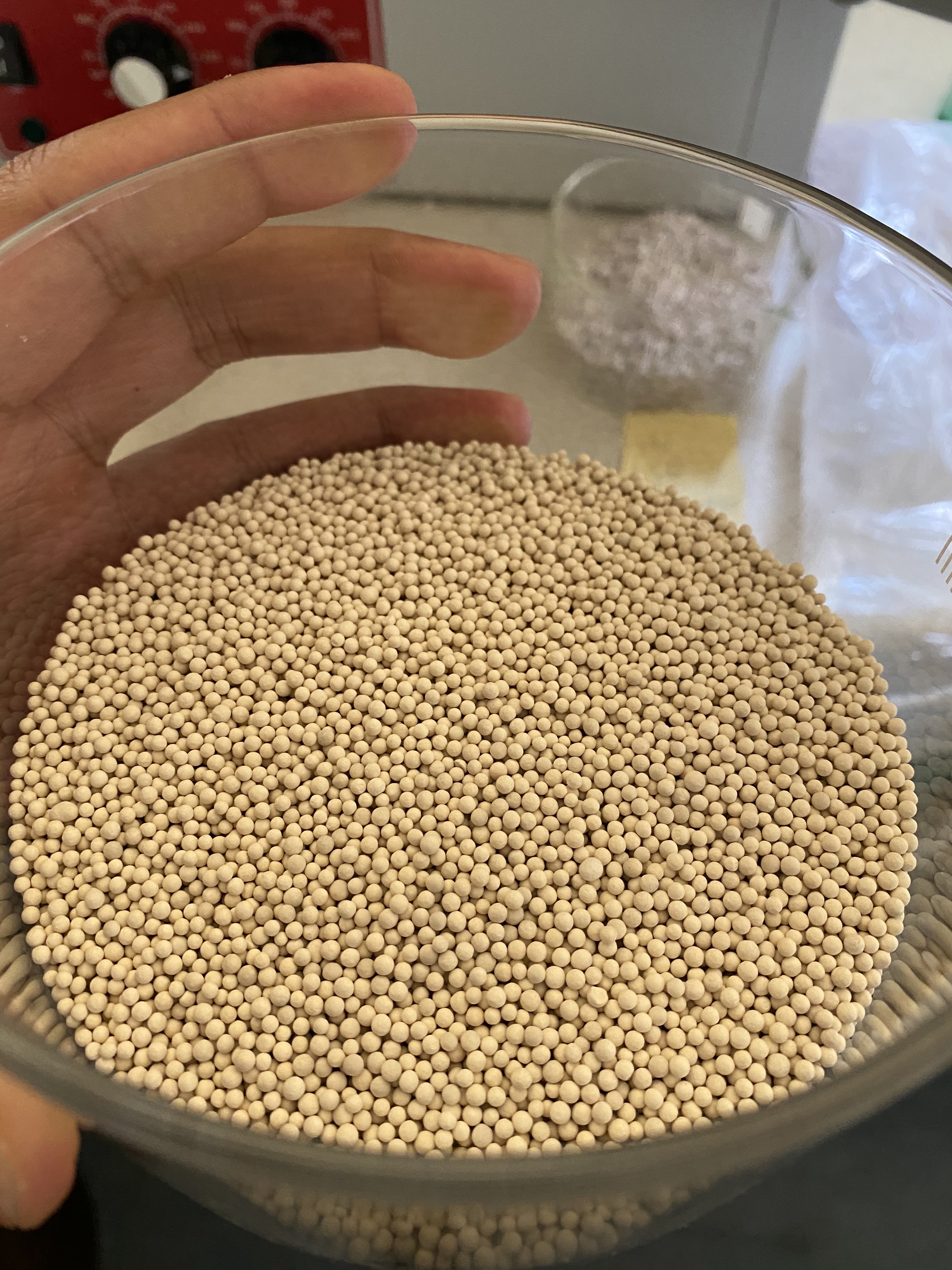}
\caption{\label{fig:MS_REAL} Images of the $5\angstrom$ type molecular sieves tested. The NU-developed MS is on the left and Sigma-Aldrich MS on the right.}
\end{figure}

Molecular sieves are chemical structures with specific pore sizes. Molecules with a critical diameter equal or smaller than the pore size will diffuse into the pores and be captured. Whereas, molecules with diameters larger than the pore size will not be captured. Instead, the larger molecules will pass between molecular sieve gaps, as shown in figure \ref{fig:MS_FILTER}. 
\\
The sizes of the pores are controlled by using different size cations in the ion exchange processes during production. In the $5\angstrom$ type molecular sieve, the cation used is calcium. The chemical formula for the $5\angstrom$ type is shown in equation \ref{eq:chemistry} \cite{sigma-aldrich_molecular_nodate}.
\begin{equation}
    \ce{0.80 CaO*0.20*Na2O *Al2O3 *2.0 SiO2*x H2O}.
    \label{eq:chemistry}
\end{equation}
The source of radioactive contamination in molecular sieves is from raw materials used in production, mainly the metallic components. In commercial production, the screening of raw materials for radioactive content is not necessary. H.Ogawa et al., from Nihon University in collaboration with Union showa K.K, has developed a method of producing low radioactive molecular sieves by extensive material selection and screening of raw materials used. The $5\angstrom$ type NU-developed MS was made by exchanging calcium ions with the low-radioactive $4\angstrom$ type MS mentioned in \cite{ogawa_development_2020}. In figure \ref{fig:MS_REAL}, the MS samples used in our comparison are shown. On the left is the $5\angstrom$ type NU-developed, in the form of white irregular granules with approximate size of 1-2 cm. On the right is the $5\angstrom$ type MS manufactured by Sigma-Aldrich in uniform beige beads with 8-12 mm diameter. 
\\
When choosing a suitable molecular sieve for a radon filtration system, it is important to understand how radon behaves in a sealed chamber. A gas-based detector is fundamentally a sealed gas chamber with various materials which can emanate radon. Once the chamber is sealed, the radon concentration inside the vessel increases continuously until a steady-state is reached. This is when the production rate of radon is equal to the decay rate of radon, termed secular equilibrium.

The steady-state radon concentration can be considered as the maximum radon activity in the chamber, applying a molecular sieve filter reduces this maximum. The determining mechanism now is the equilibration between the amount of radon captured by the MS and the amount of radon intrinsically emanated by the MS. 

To determine the suitability of a MS candidate, both the amount of radon intrinsically emanated from the sieve and its radon capturing efficiency must be measured. The ideal MS having a minimal intrinsic emanation whilst having an optimal capturing efficiency.

%% file: emanation.tex
Radon is a gaseous element produced in both the $^{238}$U and $^{232}$Th decay chains. Most materials contain a trace amount of $^{238}$U and $^{232}$Th due to inevitable material contamination, which is exacerbated by their extremely long half-lives. When radon is produced within a material, due to its gaseous nature, it can escape and emanate into its surroundings. The radon isotope $^{222}$Rn, produced from the $^{238}$U decay chain, is the most problematic isotope for contamination because $^{222}$Rn has a half-life of 3.8 days, meaning it has the ample time to reach the material surface and travel within the detector before decaying. In contrast, the second most abundant radon isotope, $^{220}$Rn, produced from the $^{232}$Th decay chain, has a half-life of 55 seconds. Hence, tests reported here focused on measurements from emanation of the $^{222}$Rn isotope.

\begin{figure}[ht]
\includegraphics[height=7cm]{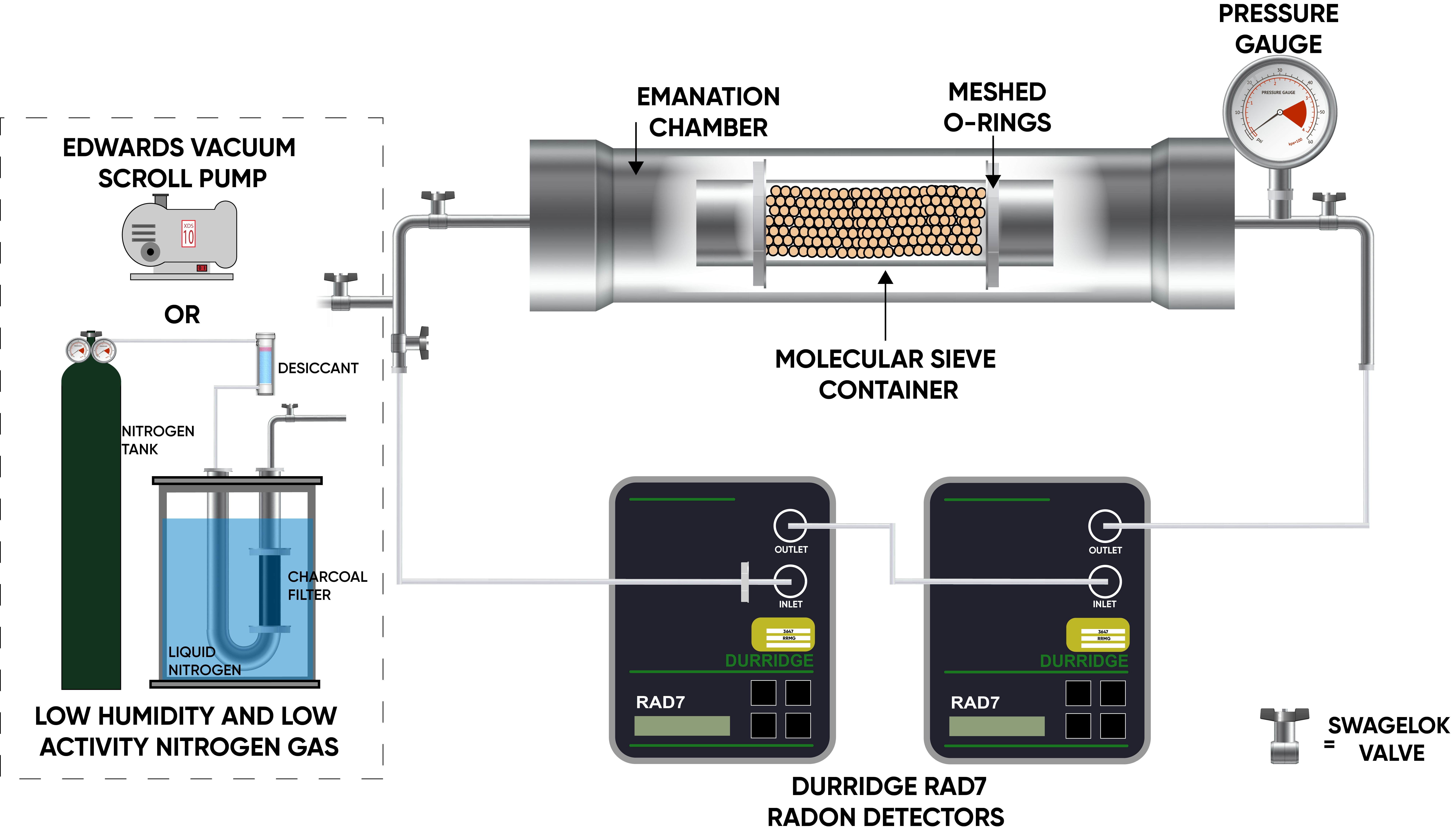}
\centering
\caption{\label{fig:emanation} Schematic of the setup used for the radon emanation tests.}
\end{figure}

\paragraph{Experimental Setup}
The radon emanation test for the molecular sieves were performed using the experimental setup shown in figure \ref{fig:emanation}. The setup consist of a 3.5L stainless steel emanation chamber in a loop with two DURRIDGE RAD7 radon detectors. The RAD7 collects radon by electrostatic precipitation, measuring alpha decays from radon progeny such as $^{218}$Po and $^{214}$Po \cite{durridge_company_inc_rad7_2020}.
The loop includes a tee connection where either an EDWARDS vacuum scroll pump or an input of low-humidity low-activity nitrogen gas was connected. The MS of interest was enclosed inside the \textit{Molecular Sieve Container} with meshed O-rings to stop any small fragments escaping from the vessel during evacuation.

\paragraph{Method}
The emanation testing procedure started with two rounds of 1-hour gas evacuations using the EDWARDS pump. The first evacuation was to get the emanation chamber into sub-torr pressures, inducing outgassing from materials. Since the initial outgassing was expected to be dominated by the emanation chamber, the gas was evacuated again after an outgassing period of 48 hours. After the second gas evacuation, the MS was left to emanate into the vacuum for seven days.\\
On completion of the emanation period, the chamber was backfilled with nitrogen gas. It was essential to use low activity-low humidity nitrogen gas to backfill the chamber. Low activity nitrogen gas was required to minimise the introduction of any ‘new’ radon to the system. Low-humid gas was required because increased humidity can suppress radon counts measured by the RAD7. Low humidity-low activity gas was achieved by flowing the nitrogen through a desiccant, which removes water molecules, and a charcoal filter cold trap, which removes any radon introduced by the nitrogen tank. Once the chamber was backfilled to 760 torr, the RAD7 detectors were purged with nitrogen for 5 minutes to remove any potential residue inside the detector. Finally, the RAD7 with its internal pump on, recorded radon concentration every four hours for a 48-hour measurement period.

\paragraph{Data Analysis}
The total radon emanating from the molecular sieves and setup is equal to the radon activity at secular equilibrium. For $^{222}$Rn to reach secular equilibrium, an emanation time in the order of a month is required. However, in our molecular sieves emanation testing, the samples were only left to emanate for seven days. To compensate for the shorter emanation time, equation \ref{eq:emanation}, which describes the accumulation of radon in a closed system as a function of emanation time was fitted to the measured data. 
\begin{equation}
    A_m=A_s \times (1-\exp{(-t/\tau)}),
    \label{eq:emanation}
\end{equation}
where $A_m$ is the measured radon activity, $A_s$ is the radon activity at secular equilibrium, $t$ is the emanation time and $\tau$ is the lifetime of $^{222}$Rn. Before the raw concentration data was fitted to equation \ref{eq:emanation}, radon concentration (Bq/m$^3$) was converted to radon activity (Bq). This was achieved by multiplying the radon concentration by the total volume in the emanation measurement loop. The total volume is given by equation \ref{eq:volume}. Where $V_{EC}$ is the volume of the emanation chamber (3.5 L), $V_{R7}$ is the volume of the RAD7 (0.9 L) and $V_{MS}$ is the volume of the molecular sieve container including the volume occupied by the molecular sieve sample (0.15 L).
\begin{equation}
    V_{T}=V_{EC}+2\times V_{R7}-V_{MS},
    \label{eq:volume}
\end{equation}
A correction to ensure there was ample radon mixing time in the measurement loop for radon to distribute evenly, was also applied. This was done by disregarding the first cycle of the RAD7 data. The nominal flow rate of the RAD7 pump was 1.0 LPM. Therefore, one measurement cycle corresponds to approximately six full system recirculation for mixing.\\
The total radon emanating from the molecular sieves was calculated by extrapolating the radon activity at secular equilibrium, $A_S$, from the equation \ref{eq:emanation} fit, and subtracting it with the background radon emanation of the setup.

\begin{figure}[ht]
\centering 
\includegraphics[width=7.5cm]{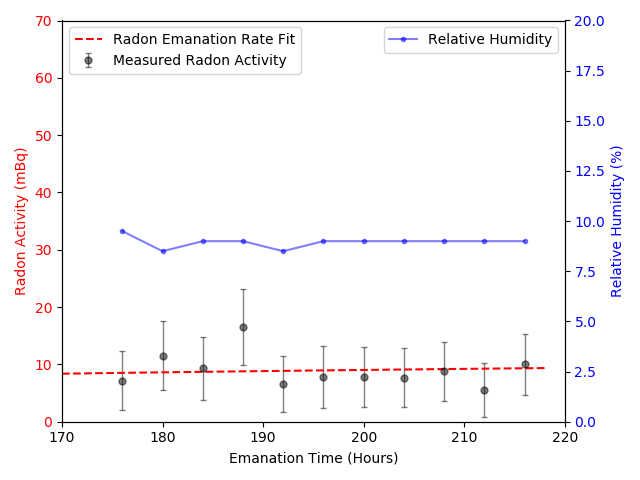}
\includegraphics[width=7.5cm]{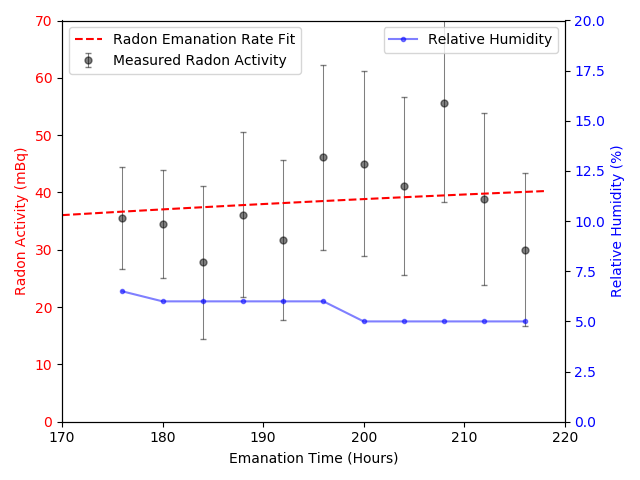}
\caption{\label{fig:emaplots} Plot of radon concentration as a function of emanation time. NU-developed MS on the left and Sigma-Aldrich MS on the right.}
\end{figure}

\paragraph{Results}
The radon concentration as a function of time is shown in figure \ref{fig:emaplots} (NU-developed MS on the left and Sigma-Aldrich MS on the right). Equation \ref{eq:emanation} non-linear regression fit are shown as dashed red lines. The extrapolated radon emanation from this fit are shown in table \ref{tab:ema}. Note that background emanation of the setup with an empty molecular sieve container was measured to be 4.69±1.1 mBq and subsequently subtracted from the results. Note that the relative humidity correction applied in section \ref{sec:filtration} was not required as the humidity was below 15\% throughout.\\
\begin{table}[ht]
\caption{\label{tab:ema} Results of the MS intrinsic radon emanation test. }
\centering
\smallskip
\begin{tabular}{|c|c|c|c|}
\hline
\begin{tabular}[c]{@{}c@{}} \textbf{Molecular} \\ \textbf{Sieve}\end{tabular} &$^{222}$\textbf{Rn Emanated (mBq)}& \begin{tabular}[c]{@{}c@{}}\textbf{Amount of} \\ \textbf{MS Used (g)}\end{tabular} & \begin{tabular}[c]{@{}c@{}} $^{222}$\textbf{Rn Emanated} \\ \textbf{per kg (mBq/kg})\end{tabular} \\
\hline
NU-developed & 6.95±1.64 & 70 & 99±23\\
Sigma-Aldrich & 45.2±3.2 & 86 &  525±37\\
\hline
\end{tabular}
\end{table}

It can be seen from table \ref{tab:ema} that the NU-developed MS has significantly lower intrinsic radon emanation per unit mass by about a factor of 5.
This result makes a promising case for the NU-developed MS to be used in radon filtration setups for ultra-low background SF$_6$-based physics experiments. However, to make a complete comparison, the radon capture efficiency of the MS must also be considered.

%% file: filtration.tex
Previously it was found that radon atoms have a critical diameter between 4 and 5$\si{\angstrom}$, while \ce{SF6} molecules have a critical diameter between 5 to 10$\si{\angstrom}$ \cite{ezeribe_demonstration_2017}. This makes it possible to remove radon atoms from \ce{SF6} with molecular sieves with 5$\si{\angstrom}$ sized pores. In the radon filtration test, the amount of $^{222}$Rn captured by the molecular sieves from radon-contaminated \ce{SF6} is measured.

\begin{figure}[ht]
\includegraphics[height=7cm]{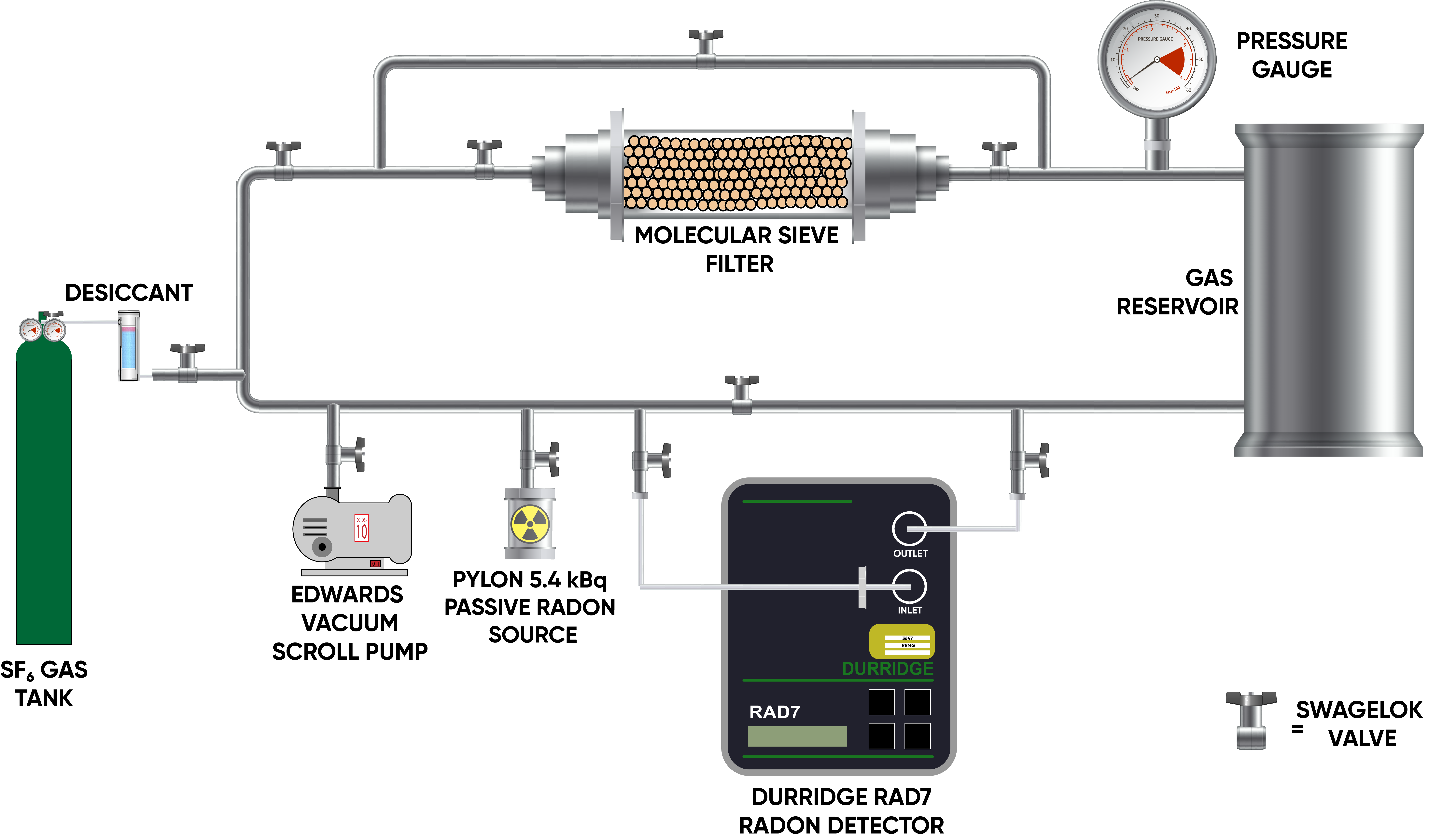}
\centering
\caption{\label{fig:filtration} Schematic of the setup used for the radon filtration tests.}
\end{figure}

\paragraph{Experimental Setup}
The radon filtration tests were performed using the experimental setup shown in figure \ref{fig:filtration}. The setup consists of the molecular sieve filter in a loop with a RAD7 radon detector and a gas reservoir. Also, included in the loop were connections to a passive PYLON 5.4kBq radon source; an SF$_6$ gas input; an EDWARDS vacuum pump and an MS filter bypass. The setup used several Swagelok valves to direct the gas flow either via the molecular sieve filter or MS bypass. Note that the RAD7 has a pump inside with a nominal flow of 1.0 litre per minute.

\newpage
\paragraph{Method}
The filtration testing procedure started with the evacuation of the setup and the introduction of low humidity SF$_6$ gas to 760 torr. The gas was then contaminated with radon by opening the valve to the PYLON radon source, which was left to diffuse for 15 minutes. The valves were initially configured so that the gas flow direction was through the MS bypass. With the pump on, the RAD7 radon detector measurement started, recording every hour. After about 24 hours, the molecular sieve filter was engaged by opening the valves at both ends of the filter and closing the MS bypass valve. The RAD7 radon detector continued to measure for at least a further 20 hours. 
\paragraph{Data Analysis}
To calculate the amount of radon captured by the MS, the data was split into \textit{MS off} and \textit{MS on} data sets, which correspond to the measured data when the gas flow was directed via the MS bypass and MS filter, respectively. Both data sets were converted from radon concentration (Bq/m$^3$) to radon activity (Bq),  by multiplying the radon concentration with the total volume in the radon filtration setup. The total volume was approximated to be the sum of the gas reservoir (6L) and the volume of the RAD7 (0.9L).\\
Two corrections were applied to the radon activity data before calculating the amount of radon captured. The first correction accounts for relative humidity (RH) in the chamber. A high concentration of water molecules in the chamber can suppress radon counts in the RAD7 detector. To compensate for lost radon counts, at RH higher than 15\%, equation \ref{eq:humidity} provided by DURRIDGE was applied to correct for high humidity. In this equation, $A_m$ is the radon activity measured, $A_c$ is the corrected radon activity and $RH$ is the associated relative humidity with the radon measurement.
\begin{equation}
    A_c=A_m \times \frac{100}{116.67-1.1\times RH}.
    \label{eq:humidity}
\end{equation}
The second correction applied ensures there was ample time for the radon from the PYLON source to distribute evenly throughout the system. Therefore, the first 4 hours of the data were disregarded. 
Using the corrected data, the radon reduction due to the application of the MS was determined by fitting the \textit{MS off} and \textit{MS on} data sets to decay equation \ref{eq:decay}. Note that the \textit{MS on} fit was optimised by only including data after a new equilibrium was reached. 
\begin{equation}
    A(t)=A_0\exp{(-\lambda t)}.
    \label{eq:decay}
\end{equation}
 $A(t)$ is the radon activity at time $t$, $A_0$ is the initial radon activity and $\lambda$ is the radon decay constant. The initial radon activity, $A_0$, for both \textit{MS off} and \textit{MS on} data sets were extrapolated, and the effective radon reduction was determined by calculating the difference between the extrapolated $A_0$ values.

\begin{figure}[ht]
\centering 
\includegraphics[width=7cm]{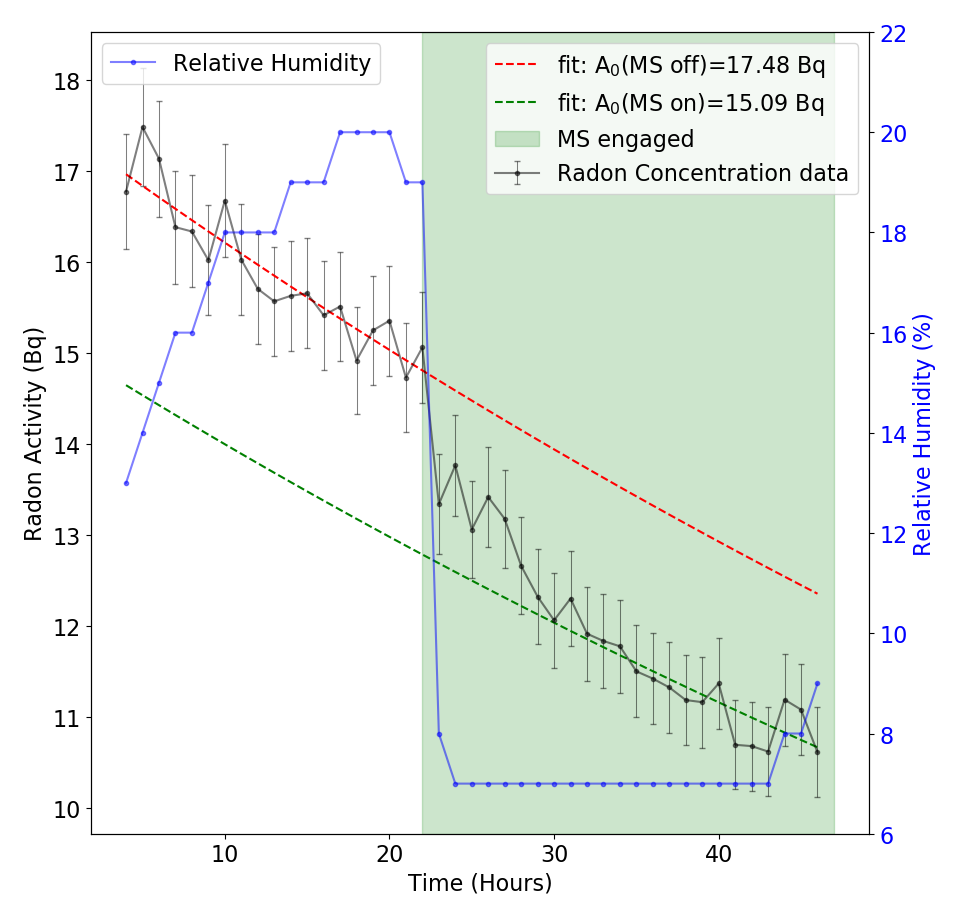}
\includegraphics[width=7cm]{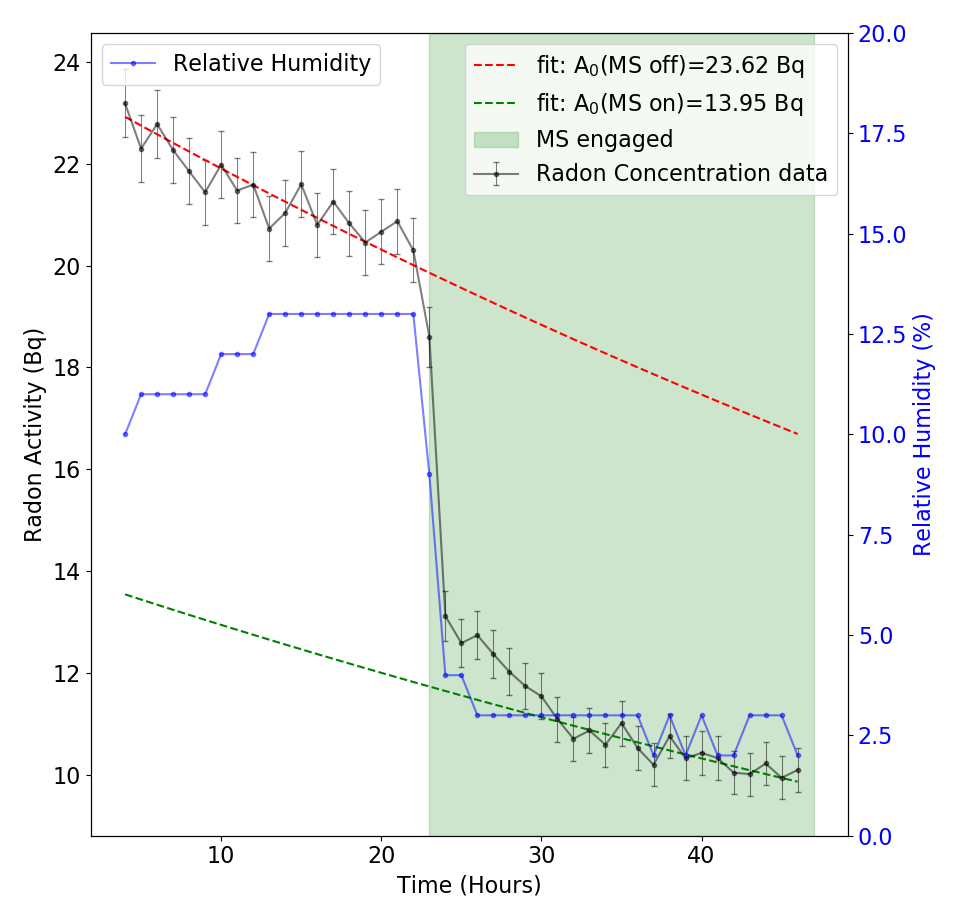}
\caption{\label{fig:filtresult} Plot of the change in radon concentration observed due to application of molecular sieve. NU-developed MS on the left and Sigma-Aldrich MS on the right.}
\end{figure}

\paragraph{Results} 
The radon activity as a function of time is shown in figure \ref{fig:filtresult} (NU-developed MS on the left and Sigma-Aldrich MS on the right). The \textit{MS off} and \textit{MS on} non-linear regression fits are shown in red and green dashed lines, respectively. The highlighted green area represents that the molecular sieve filter was engaged. The blue points represents the relative humidity in the setup. 
The implementation of the molecular sieves, resulted in a drop in relative humidity, as expected. In addition, it is clear that the effective radon activity also decreases, results shown in table \ref{tab:filt}. The NU-developed MS demonstrates the removal of radon from SF$_6$. However, at a lower radon capture efficiency(35±2 Bq kg$^{-1}$) compared to the Sigma-Aldrich MS (97±1 Bq kg$^{-1}$).\\
\begin{table}[ht]
\caption{\label{tab:filt} Results of the radon filtration test.}
\centering
\smallskip
\begin{tabular}{|c|c|c|c|}
\hline
\begin{tabular}[c]{@{}c@{}} \textbf{Molecular} \\ \textbf{Sieve}\end{tabular} &\textbf{Rn Captured (Bq)} & \begin{tabular}[c]{@{}c@{}}\textbf{Amount of} \\ \textbf{MS Used (g)}\end{tabular} & \begin{tabular}[c]{@{}c@{}} $^{222}$\textbf{Rn Captured} \\ \textbf{ per kg (Bq kg$^{-1}$})\end{tabular} \\
\hline
NU-developed & 2.39±0.10 & 68& 35±2 \\
Sigma-Aldrich & 9.67±0.12 & 100 & 97±1\\
\hline
\end{tabular}
\end{table}

%% file: comparison.tex
To provide a complete comparison of the molecular sieve candidates, the results from the emanation and filter tests were combined. A parameter indicating the amount of radon emanated by the MS per radon captured is shown in table \ref{tab:comparisonfinal}. This parameter was used as the figure of merit for the suitability of the MS for use in radon filtration setups for ultra-low background SF$_6$-based physics experiments, with a value that needs to be minimised. The NU-developed $5\angstrom$ MS emanated radon 48±15$\%$ less per radon captured, compared to the commercial Sigma-Aldrich MS. This result shows that the NU-developed MS is a better candidate.
\begin{table}[ht]
\caption{\label{tab:comparisonfinal} Calculated comparison parameter, indicating the amount of radon emanated by the MS per radon captured by the MS from \ce{SF6}. }
\centering
\begin{tabular}{|c|c|}
\hline
\textbf{Molecular Sieve} &
\begin{tabular}[c]{@{}c@{}} $^{222}$\textbf{Rn Emanated per}\\ \textbf{ $^{222}$\textbf{Rn captured} ($\times 10^{-3}$)}\end{tabular} \\
\hline
NU-developed &  2.8±0.7\\
Sigma-Aldrich  & 5.4±0.4\\
\hline
\end{tabular}
\end{table}

\noindent To determine if the NU-developed MS activity levels are acceptable for ultra-low background SF$_6$ gas-based physics experiments, let us consider a practical example. The DRIFT (Directional Recoil Identification From Tracks) Experiment is a gas-based direction dark matter experiment with chamber dimensions of 1.5 x 1.5 x 1.5 m$^3$. The radon activity due to material contamination in the experiment was measured to be 372±66 mBq \cite{battat_radon_2014}.The 1 mBq radon background target, allows for up to 13g of NU-developed MS granules to be used. This amount equates to a capacity of 455±26 mBq, which is sufficient to capture all the radon in the detector (372±66 mBq). However, since radon capture is a probabilistic kinetic process, using a small amount of MS to capture a limited amount of radon atoms in a large volume would make the timescales problematic. Therefore, the filtration rate must be optimised when designing the gas system, maximising the contact time between the MS and the gas. Equally, using more sieves would improve the filtration rate as more pores will be available to capture radon. As larger-scale experiments will have more materials, radon contamination levels will inevitably be higher. Therefore, efforts must continue towards minimising the emanation per radon captured parameter so that the total amount of MS allowed by the radioactive budget of an experiment is maximised.

%% file: geometry.tex
The radon reduction result for the NU-developed MS is significantly less than the Sigma-Aldrich MS result, seen in table \ref{tab:filt}. Optimising the radon filtration for NU-developed MS could possibly improve the radon emanated per radon captured parameter. The filtration discrepancy may be attributed to the difference in their geometries and surface area to volume ratio, seen in figure \ref{fig:MS_REAL}.

\begin{figure}[ht]
\centering 
\includegraphics[height=3cm]{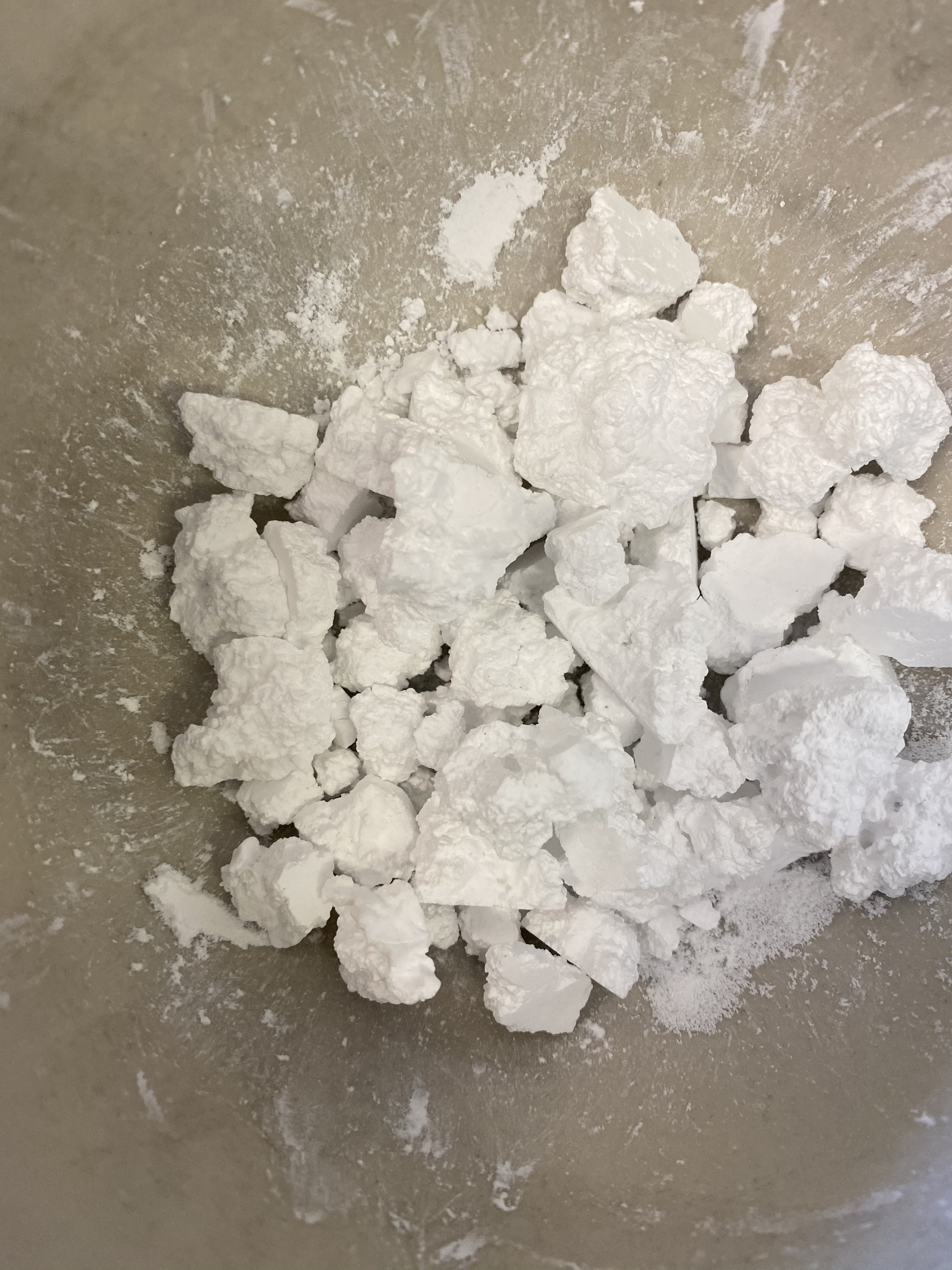}
\qquad
\includegraphics[height=3cm]{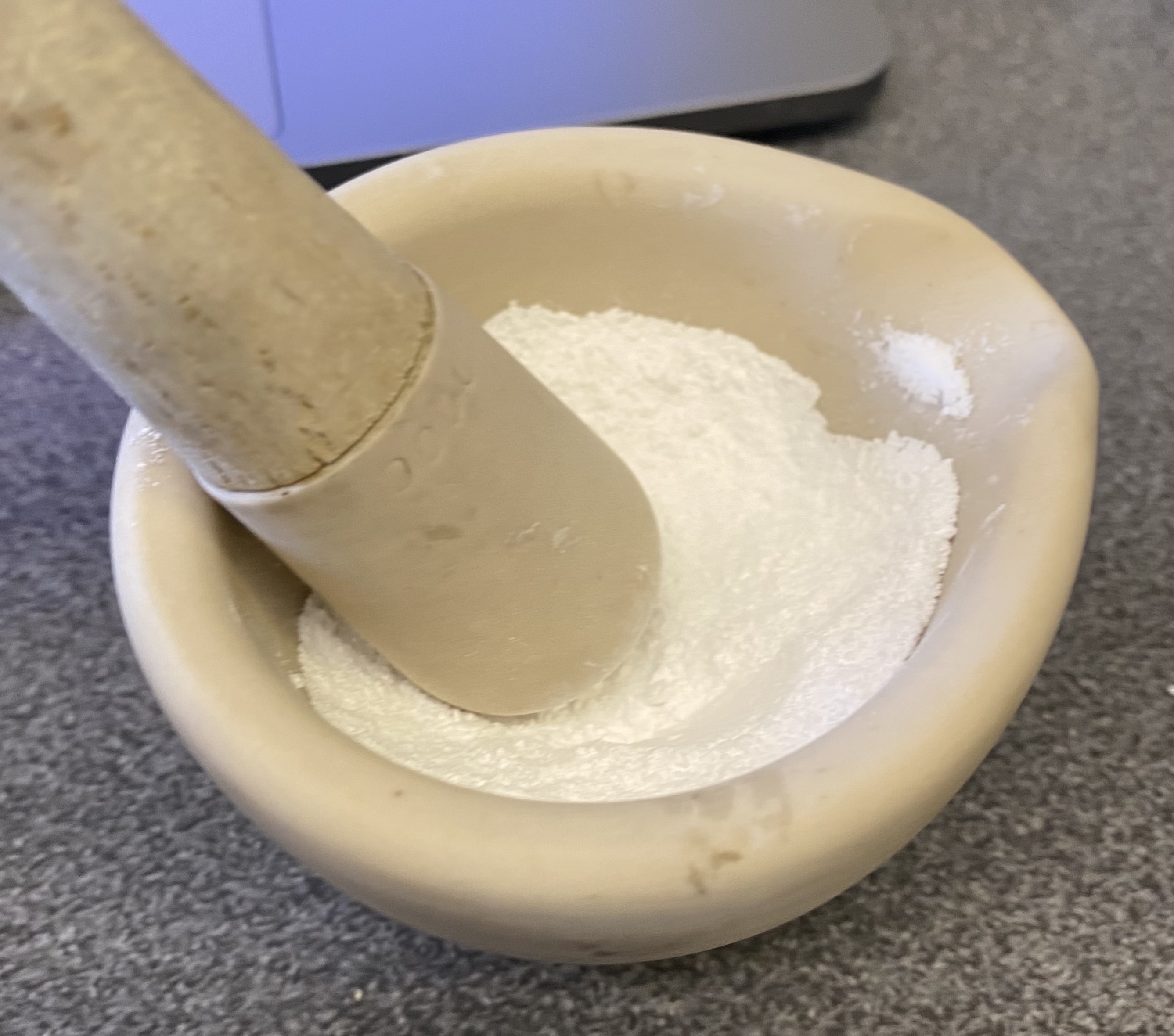}
\caption{\label{fig:powder} Images of the NU-developed MS before and after crushing it into a fine powder.}
\end{figure}

\begin{figure}[ht]
\centering 
\includegraphics[width=7.5cm]{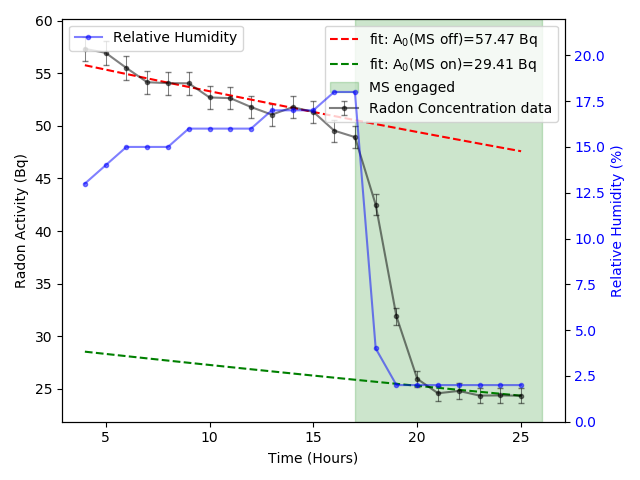}
\includegraphics[width=7.5cm]{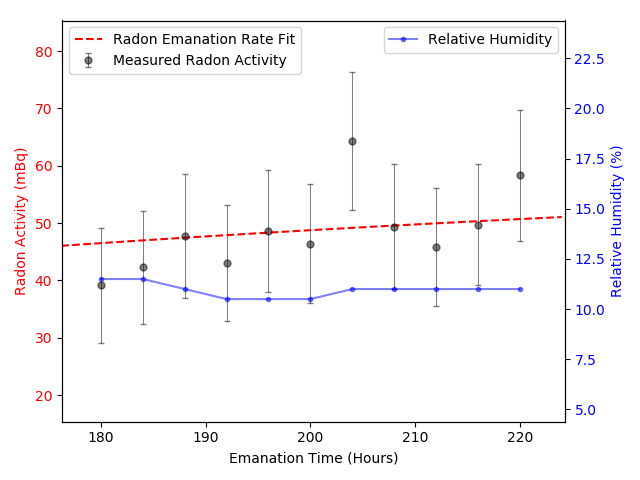}
\caption{\label{fig:georesult} Plot of the change in radon concentration observed while using the powdered NU-developed MS (left) and a plot of radon concentration as a function of emanation time (right) in the powdered NU-developed MS emanation test. }
\end{figure}

In a bid to improve the radon capturing capabilities of the NU-developed MS, its surface area to volume ratio was increased by crushing it from its irregular granular form was into a fine powder with a mortar and pestle, as shown in figure \ref{fig:powder}. The radon filtration test, discussed in section \ref{sec:filtration}, was repeated with the NU-developed MS in its powder form. It is worth noting that the significant increase in surface area to volume ratio has implications that there will be an increase in the amount of radon emanated form the MS, as the path from inside the material to the surface is reduced. Therefore, the emanation test, discussed in section \ref{sec:emanation}, was also repeated with the powder form.
 \\
The results for the powder form of the NU-developed MS test are shown in table \ref{tab:geometryfilt}, with the corresponding filtration and emanation plots in figure \ref{fig:georesult}. The radon reduction efficiency has significantly improved from 35±2 to 330±3 Bq kg$^{-1}$. As expected, the intrinsic radon emanation has significantly worsen from 99±23 to 680±30 mBq. However, the radon emanated per radon captured parameter has remained the same within errors. Overall, crushing the MS into powder did not improve is suitability for use in radon filtration setups for ultra-low background as demonstrated by our parameter.

\begin{table}[ht]
\caption{\label{tab:geometryfilt} Radon filtration, intrinsic MS emanation and comparison parameter results for the NU-developed MS in granule and powdered form. Note that 85g of NU-developed MS were used in the powdered tests.}
\centering
\smallskip
\begin{tabular}{|c|c|c|c|}
\hline
\begin{tabular}[c]{@{}c@{}} \textbf{NU-developed} \\ \textbf{MS}\end{tabular} 
&\begin{tabular}[c]{@{}c@{}} $^{222}$\textbf{Rn Captured} \\ \textbf{ per kg (Bq kg$^{-1}$})\end{tabular}  & 
\begin{tabular}[c]{@{}c@{}} $^{222}$\textbf{Rn Emanated} \\ \textbf{ per kg (mBq kg$^{-1}$})\end{tabular}  &
\begin{tabular}[c]{@{}c@{}} $^{222}$\textbf{Rn Emanated per} \\ \ $^{222}$\textbf{Rn Captured ($\times 10^{-3}$)}\end{tabular} \\
\hline
Granules  & 35±2 & 99±23 & 2.8±0.7 \\
Powder & 330±3 &  680±30 & 2.1±0.1 \\
\hline
\end{tabular}
\end{table}